\documentclass[12pt, a4paper]{article}

\newcommand{\be}{\begin{equation}}
\newcommand{\ee}{\end{equation}}
\newcommand{\bea}{\begin{eqnarray}}
\newcommand{\eea}{\end{eqnarray}}
\newcommand{\no}{\noindent}

\title{\sc One dimensional M5-brane intersections}
\author{Ansar Fayyazuddin$^{1,2}$\footnote{email: ansar@physto.se},
Tasneem Zehra Husain$^{2,3}$\footnote{email: tasneem@physto.se}\\
and {Dileep P. Jatkar}$^{3,4}$\footnote{email: dileep@mri.ernet.in}}

\begin{document}

\date{}

\maketitle

\begin{center}

{\em \vspace{-0.3cm}
$^1$Department of Physics and Astronomy\\
Tufts University, Medford, MA 02155,
USA\\
\vspace{0.2cm}
$^2$Department of Physics, Stockholm University\\
Box 6730, S-113 85 Stockholm,
Sweden \\
\vspace{0.2cm}
$^3$Jefferson Physical Laboratory,
Harvard University,\\
Cambridge, MA
USA \\
\vspace{0.2cm}
$^4$Harish-Chandra Research Institute\\
Chhatnag Road, Jhusi, Allahabad, 211019 India}

\end{center}

\vspace{-0.2cm}

\begin{abstract}
We study one dimensional intersections of M5 branes with M5 and M2 branes. On
the worldvolume of the M5-brane, such an intersection appears as a string
soliton. We study this worldvolume theory in two different regimes:  1) Where
the worldvolume theory is formulated in flat space and 2) where the
worldvolume theory is studied in the supergravity background produced by a
stack of M5 (or M2) branes. In both cases, we study the corresponding string
solitons, and find the most general BPS configuration consistent with the
fraction of supersymmetries preserved.  We argue that M5 and M2 brane
intersections leave different imprints on the worldvolume theory of the
intersecting probe brane, although geometrically they appear to be
similar.

\end{abstract}
\vspace{-21cm}
\begin{flushright}
USITP-04-05 \\
hep-th/0407129
\end{flushright}

\thispagestyle{empty}

\newpage

\setcounter{page}{1}

\section{Introduction}

The one dimensional intersection of M5-branes is a much neglected
subject. One of the reasons this system is not better understood is
simply that it does not obey the $(p - 2)$-rule\cite{Tseytlin} which
generates all other self-intersecting M-brane
configuration\cite{Itzhaki:1998uz}--\cite{LastPaper}.

This rule\cite{Tseytlin} states that one $p$-brane can intersect another
along $(p-2)$ spatial directions, if the resulting system is required to be
BPS. By extension, if each pair of branes in a particular configuration
intersects along $(p - 2)$ spatial directions, we are guaranteed that the
resulting configurations preserves some supersymmetry. Due to its simplicity
and vast jurisdiction, this rule has been used extensively to write down
complicated BPS configurations.

The defiance of the $(p - 2)$ rule by the M5$\perp$M5(1) configuration, was
thus, both a mystery and an obstacle to a clear understanding of the system.
However, it was claimed in \cite{braneteaser} that this particular system is
exempt from the rule! The reasoning is as follows: It turns out that the
presence of any worldvolume fields (other than scalars and their duals)
results in a contradiction of the assumptions under which the $(p - 2)$
rule was `derived'. Since the M5$\perp$M5(1) system is the only
self-intersecting brane configuration for which the world-volume two-form on
the fivebrane is turned on, that rule does not apply to this configuration.

Encouraged by the fact that some inroads are finally being made into
understanding this system, we turn our attention here to another unsolved
issue regarding M5$\perp$M5(1). In \cite{gaunt} it is claimed that whenever
M5-branes intersect over one dimension an M2-brane is always secretly
present. In this paper, we try to find whether or not that claim indeed holds
true. To this end, we look at one dimensional solitons on M5-branes in
several ways; our analysis will thus be confined to the worldvolume of a
fivebrane.

We will start by considering an M5 worldvolume theory in flat space and looking
for string solitons in that theory. We display the most general solution for
the soliton preserving the appropriate amount of supersymmetries, and proceed
to study an M5-brane theory in the supergravity background produced by,
alternately, infinite M5 and M2 branes.

We will show that the induced metric on the M5-brane is different in the two
cases. Although, in the near-horizon Maldacena limit of the background branes
the geometry is $AdS_3\times S^3$ for both backgrounds, the radii of curvature
distinguish the two cases. In fact, these radii of curvature match only for
very special values of the number of branes.  We take this as evidence that the
one dimensional M5/M5-brane intersection is a genuine intersection and not just
a secret M2 brane stretching between the two M5 branes.

It is perhaps worth mentioning at this point, that even in string
theory, the lower dimensional descendants of the enigmatic
M5$\perp$M5(1) system continue to be somewhat special and are
definitely not completely understood. The one dimensional
intersection of two M5-branes can be dimensionally reduced to two
D4-branes intersecting over a point. This system is known to have
two supersymmetric branches in the presence of a B-field; one of
which branches is continuously connected to $B=0$, whereas the other
is not \cite{park,Witten} T-dualizing further relates it to an
intersecting D0-D8 system, which again, is interesting in its own
right.

\section{String solitons on M5 Worldvolume}

\subsection{Cause for confusion}

Consider two M5-branes, one of them extended in the 12345 directions and the
other along 16789. Using $\times$ to denote directions tangent to the
world-volume of a brane, we can express this configuration in tabular form as
follows:

\begin{equation}
\begin{array}[h]{|c|cc|cccc|cccc|c|}
  \hline
   \; & 0 & 1 & 2 & 3 &
              4 & 5 & 6 & 7 & 8 & 9 & 10\\
  \hline
  {\bf M5} & \times & \times & \times & \times & \times & \times
               &  &  &  &  &  \\
  {\it M5} & \times & \times
               &  &  &  &  & \times & \times & \times & \times  & \\
  \hline
\end{array}
\label{m5m5}
\end{equation}

The two M5-branes can be separated along $X^{10}$. If, however, they are
located at the same point in $X^{10}$, they must intersect along $X^1$. We can
study this system from the point of view of the worldvolume of one of the
M5-branes, which we denote as {\bf M5}.  The other M5-brane\footnote{Throughout
this paper we use boldface notation for the probe brane, and italics for
background branes.}, denoted by {\it M5}, will then appear as a soliton in the
worldvolume theory of {\bf M5}.

Similarly, one can introduce an {\it M2}-brane stretched along $X^{10}$ and
ending on the {\bf M5}-brane along $X^1$.

\begin{equation}
\begin{array}[h]{|c|cc|cccc|cccc|c|}
  \hline
   \; & 0 & 1 & 2 & 3 &
              4 & 5 & 6 & 7 & 8 & 9 & 10\\
  \hline
  {\bf M5} & \times & \times & \times & \times & \times & \times
               &  &  &  &  &  \\
  {\it M2} & \times & \times
               &  &  &  &  &  &  &  &  & \times \\
  \hline
\end{array}
\label{m5m2}
\end{equation}

The one dimensional end of this M2-brane will again appear as a worldvolume
soliton in the M5-brane.

{}From the bulk vantage point, these two configurations are very similar -
perhaps even confusingly so. In both cases the surrounding spacetime is static
and exhibits translation symmetry along $X^1$.  Both configurations have an
$SO(4) \times SO(4)$ isometry corresponding to rotational symmetry in the $X^2
\cdots X^5$ and $X^6 \cdots X^9$ and $X^{10}$.

In addition, these two intersecting brane systems require identical projection
conditions to be satisfied by Killing spinors, and thus preserve the same
amount of supersymmetry. Killing spinors for the $M5 \perp M5$ configuration
described in ({\ref{m5m5}}), obey
\begin{eqnarray}
\gamma_{012345} \epsilon &=& \epsilon \nonumber \\
\gamma_{016789} \epsilon &=& \epsilon
\end{eqnarray}
whereas the Killing spinors for the $M5 \perp M2$ system of ({\ref{m5m2}}) are
such that
\begin{eqnarray}
\gamma_{012345} \epsilon = \epsilon \\
\gamma_{01(10)} \epsilon = \epsilon
\end{eqnarray}
Using the identity $\gamma_{(10)} = \gamma_0 \gamma_1 \cdots \gamma_9$, it is
trivial to see that the two sets of constraints given above are in fact
identical; any one set implies the other.

Since we are currently limiting ourselves to looking at the system in its probe
approximation, the bending of spacetime due to any of the above M-branes need
not be taken into account; in this section, the background is thus considered
to be flat.

Given the fact that the one dimensional intersection of an M5 brane with
M5-branes is similar in so many ways to the one-dimensional ending of a
membrane on it, it is important to ask how these two configurations can be
distinguished; could they perhaps have different manifestations on the
worldvolume of {\bf M5}? It is to answer this question that we investigate how
{\it M2} and {\it M5} appear, when viewed from the point of view of (the probe)
{\bf M5}.  As we will show, in the flat background, for every integer value of
the charge $Q$ with respect to the self dual three form field strength $\hat H$
on the {\bf M5} brane worldvolume, there exists a unique BPS soliton which
respects the isometries required by a 1-dimensional intersection. In this
situation, the world volume string soliton will not be able to distinguish {\it
M2} ending on {\bf M5} from M5$\perp$M5(1). However, in the next section, we
will show that there is an interesting twist to this story when we take into
account curvature effects due to background {\it M2/M5} branes.

\subsection{The View from the WorldVolume}

In the next subsection we will obtain the BPS string soliton on the {\bf M5}
brane world volume in the flat background, however, in this subsection we will
set up BPS conditions for the soliton without making any specific choice of
background metric. The worldvolume theory we will study is that of an M5-brane
oriented along 12345.  The bosonic content of this theory consists of five
scalars ($X^a, a=6,..10$) corresponding to the transverse fluctuations of the
M5-brane and a 2-form whose associated field strength is self-dual.

Let us now consider a one dimensional soliton oriented along, say, the $X^1$
direction. As discussed previously, there is an SO(4) symmetry in the remaining
world-volume directions, $X^2,\cdots ,X^5$. If we define a radial coordinate
$\tilde r^2=\sum_{i=2}^{5}(X^i)^2$, this isometry is reflected in the fact that
scalar fields depend only on $\tilde r$.

The scalar fields $X^a, a=6,\cdots ,10$ can be split into two groups: one
consisting of $X^6,\cdots ,X^9$ which are scalars representing directions along
the {\it M5}-brane soliton (or transverse to the {\it M2}-brane), and $X^{10}$
which is transverse to the {\it M5}-brane soliton (or along the {\it
M2}-brane).

Since we are looking for a string soliton, we expect the two-form to have
components along the $0,1$ directions and to depend only on $\tilde r$.  The
self-duality of the corresponding three-form field strength implies that we
have not only the components $\hat H_{01\tilde r}$, but also its dual
components along the transverse $S^3$. For ease of notation we now define a
`reduced two-form' $H_{ab} \equiv \hat H_{0ab}/\sqrt{g_{00}}$.

In order for a configuration to be supersymmetric, it must saturate
a BPS bound, the worldvolume formulation of which for flat space is
given in \cite{sorokin, m2onm5}. This can be written in a form which
generalizes to curved space:
\begin{equation}\label{bpsc}
\sqrt{\det(g+H)} = \epsilon^{\dagger}\gamma^0\left[\frac{1}{5!}
\Gamma_{abcde}\epsilon^{abcde}-\frac{\sqrt{\det(g)}}{2}\Gamma_{ab}
H^{ab} + \Gamma_at^a\right]\epsilon,
\end{equation}
Here, $g_{ab}$ and $\Gamma$ denote the world-volume pull-backs of the
space-time metric and $\gamma$-matrices. In static gauge we have the
expressions
\begin{eqnarray}
\Gamma_a = \gamma_a + \partial_a X^i \gamma_i \nonumber \\
g_{ab} = h_{ab} + \partial_a X^i \partial_b X^j h_{ij}
\end{eqnarray}
where $a = 0, \cdots 5$ is a worldvolume index, $i = 6 \cdots 10$ labels
directions transverse to the probe {\bf M5} brane, and $(h_{ab}, h_{ij})$
comprises the full metric in spacetime. In addition to solving the BPS
saturation condition, the field strength $H_{ab}$ should also satisfy the
Bianchi identity constraint corresponding to the gauge invariance of the two
form potential.  Using the isometries of our solution this condition can be
expressed as
\begin{equation}\label{HBI}
\partial_{\tilde r} (\sqrt{g} H^{1\tilde r}) = 0.
\end{equation}

\subsection{The string soliton solution}

As mentioned in the previous subsection, here we will restrict ourselves to the
case of flat metric in the eleven dimensional spacetime. In a flat background,
the pullback of the space-time metric onto the worldvolume (012345) of
the {\bf M5} brane is
\begin{equation}\label{wvmet}
ds^2_6 = -dt^2+dX_1^2+\left(1+\sum_{i=6}^{9}(\partial_{\tilde
r}X_i)^2 +(\partial_{\tilde r}X_{10})^2\right)d\tilde r^2 +\tilde
r^2 d\Omega_{s^3}^2.
\end{equation}

Given this induced metric, the general form of $H^{ab}$, and the spinor
projection conditions in eq.(\ref{bpsc}), we can use the BPS condition to
determine the functional dependence of $H$ on the transverse scalars. A
general solution is
\begin{eqnarray}\label{genH}
H_{1\tilde r} &=& \sqrt{1+\sum_{k=6}^{9}(\partial_{\tilde r}X_k)^2
+(\partial_{\tilde r}X_{10})^2}\times\nonumber \\
&&\left[-\partial_{\tilde r}X_{10} \pm
\sqrt{\left\{-1-\sum_{i=6}^{9}(\partial_{\tilde r}X_i)^2
-(\partial_{\tilde
r}X_{10})^2\right\}\sum_{j=6}^{9}(\partial_{\tilde r}X_j)^2}\right].
\end{eqnarray}

\no
The field strength is real if and only if we set
\begin{equation}\label{setz}
\sum_{j=6}^{9}(\partial_{\tilde r}X_j)^2 =0,
\end{equation}
which in turn means
\begin{equation}\label{realH}
H_{1\tilde r} = -(\partial_{\tilde r}X_{10})\sqrt{1
+(\partial_{\tilde r}X_{10})^2}.
\end{equation}

Notice that for the calculations we carried out in this section, we never had
to state explicitly whether we were considering the $M5 \perp M5(1)$
configuration of (\ref{m5m5}), or the $M5 \perp M2(1)$ configuration of
(\ref{m5m2}). All that we needed in order to solve the BPS equation was a
knowledge of the symmetries on the worldvolume (which dictate the form of H),
and the projection conditions on the Killing spinor. Since the preserved
supersymmetries and isometries of both (\ref{m5m5}) and (\ref{m5m2}) are
identical, it is not within the scope of our current calculation to distinguish
between these two scenarios; the results we have obtained thus far are hence
equally valid whether the background contains an M5 brane in 16789 directions
or an M2 brane in the 1(10) directions.

As we have shown, the only scalar field we can turn on while preserving
spherical symmetry and worldvolume supersymmetry is the $X^{10}$ field.  The
other scalar fields remain constant.  This solution is subject to the Bianchi
identity constraint \cite{sorokin,m2onm5} generalized to general
curved space background:
\begin{equation}\label{bic}
\partial_{\tilde r} (\sqrt{g} H^{1\tilde r}) = 0.
\end{equation}
Writing this condition in term of $X_{10}$ implies it satisfies the
equation
\begin{equation}\label{poi}
\partial_{\tilde r}(\tilde r^3\partial_{\tilde r}X_{10}) =0.
\end{equation}
or
\begin{equation}\label{soln}
X_{10} = \mbox{const.} + q_0/r^2,
\end{equation}
where, $q_0$ is proportional to the soliton number $N_1$.

\section{Probing M2/M5 backgrounds}

In this section we study the worldvolume theories of an {\bf M5}-brane probe in
the background geometries produced, alternately, by an {\it M2}-brane and an
{\it M5}-brane. We will start with the gravitational background of an {\it M2}
brane extended in the spatial directions $x^1$ and $x^{10}$. Our probe {\bf M5}
brane is extended, as usual, in the 12345 directions.

\subsection{M2-brane background}
The metric due to the {\it M2} brane background is
\begin{equation}\label{m2back}
ds^2 =
h_2^{1/3}(r)\left[h_2^{-1}(r)(-dt^2+dx_1^2+dx_{10}^2)+\sum_{i=2}^9
dx_i^2\right].
\end{equation}
The pullback of this background on the {\bf M5} brane is
\begin{eqnarray}\label{m2onm5}
ds_6^2 &=& h_2^{-2/3}(r)(-{dt}^2 + {dx^1}^2) +h_2^{1/3}(r)
\tilde r^2(d\Omega_{S^3}^2)\nonumber \\
&+&h_2^{1/3}(r)(1+\sum_{i=6}^{9} (\partial_{\tilde r}X^i)^2 +
h_2^{-1}(r)(\partial_{\tilde r}X^{10})^2) d\tilde r^2,
\end{eqnarray}
where, $h_2(r) = 1+ \frac{k}{r^6}$, $r^2 =\sum_{I=2}^{9} (x^I)^2$ and $\tilde
r^2 = \sum_{i=2}^{5}(x^i)^2$. The parameter $k=2^5\pi^2 N_2\ell_p^6$
\cite{Maldacena} depends on $N_2$, the number of {\it M2} branes, and,
$\ell_p$, the eleven dimensional Planck length.

It is simple to see (and the general results below contain it as a sub-case)
that this is a BPS solution with $H_{ab}=0$ and the $X^i=0$ , $X_{10} = 0$ in
the worldvolume theory.  This is, of course, not surprising since the M5 brane
is probing a stack of $N_2$ M2 branes which are infinite and therefore pass
right through the probe with all the charge canceling locally.  All the scalars
are constants since there is no bending due to the tension of the M2-brane
since the force due to it is locally cancelled.

It is generally difficult to compare different geometries (metrics) since they
are expressed in different coordinates.  It is, therefore, instructive to take
the Maldacena (near horizon) limit where we express the metric in the rescaled
variable $u = r^2/\ell_p^3$ \cite{Maldacena} in the low-energy limit $u\ell_p
<<1$.   In this limit, with the scalars set to constants and $H_{ab}=0$, the
geometry becomes $AdS_3\times S^{3}$.  With the AdS$_3$ radius of curvature
$R_{AdS_3}^2= \ell_p^2(\frac{\pi^2N_2}{2})^{1/3}$ and $R^2_{S^3} =
\ell_p^2(2^5\pi^2N_2)^{1/3}$\footnote{This result is complementary to
the relation obtained earlier\cite{berman} in a related context.}. This
can be compared to the M5-brane case below.

\subsection{M5-brane background}

It is instructive to contrast this situation with the curved geometry generated
by an {\it M5} brane. To be able to directly compare the two situations we will
assume the background {\it M5} brane is extended in 16789 direction and the
probe {\bf M5} brane is still extended in 12345 directions. Like in the
previous case, we will start with the background metric generated by the {\it
M5} brane extended in 16789 direction
\begin{equation}\label{m5back}
ds^2 = h_5^{2/3}(r)\left[h_5^{-1}(r)(-dt^2+dx_1^2+\sum_{i=6}^{9}dx_i^2)
+ \sum_{a=2}^{5} dx_a^2 + dx_{10}^2\right].
\end{equation}
The pullback of this metric on the probe gives the induced
worldvolume metric on M5 brane extended in 12345 direction
\begin{eqnarray}\label{m5onm5}
ds_6^2 &=& h_5^{-1/3}(r)(-{dt}^2 + {dx^1}^2) +h_5^{2/3}(r)
\tilde r^2(d\Omega_{S^3}^2)\nonumber \\
&+&h_5^{2/3}(r)(1+h_5^{-1}(r)\sum_{i=6}^{9} (\partial_{\tilde r}X^i)^2 +
(\partial_{\tilde r}X^{10})^2) d\tilde r^2,
\end{eqnarray}
where, $h_5(r) = 1+ q/r^3$, $\tilde r$ is as defined earlier, $r^2 =
\tilde r^2 + x_{10}^2$ and $q=\pi N_5\ell_p^3$ depends on $N_5$, the
number of {\it M5} branes.

Due to the different functional dependence of the harmonic function $h(r)$ in
{\it M2} and {\it M5} case, it may appear that these are two totally different
metrics. It is, therefore, useful to take the near horizon limit to see that in
both cases the near horizon geometry is $AdS_{3}\times S^3$.  The formal
similarity of these two backgrounds, however, ends here. To get the metric in
the desired form we need to take a different limit \cite{Maldacena}, with $u^2
= r/\ell_p^3$.  The radii of curvature of $S^3$ and that of $AdS_3$ are given
by $R_{S^3}^2=R_{AdS_3}^2/4 = (\pi N_5)^{1/3}\ell_p^2$. It is easy to see that
these two geometries will not be the same for arbitrary integer values for
$N_2$ and $N_5$.

\subsection{String soliton in curved background}

We will now turn our attention to the BPS string soliton, which is
extended along the $x_1$ direction, on the worldvolume of {\bf M5} brane
in the {\it M2} brane background. As in the previous section we again
take a radial ansatz for the soliton on the worldvolume and define
the radial direction by $\tilde r = \sqrt{x_2^2+\cdots + x_5^2}$.
The BPS condition implies the induced metric and the three form
field strength field $H$ should satisfy (\ref{bpsc}). We substitute
the eq.(\ref{m2onm5}) into the BPS condition (\ref{bpsc}) to
determine
\begin{equation}\label{orighm2}
\tilde H_{1\tilde r} = - \frac{h_2^{-1/3}(r)A^2(r)\partial_{\tilde
r}X^{10} \pm B(r)}{h_2^{1/3}(r)A(r)(1+\sum_{i=6}^{9}
(\partial_{\tilde r}X^i)^2)},
\end{equation}
where,
\begin{eqnarray}
A^2(r)&=& \left(1+\sum_{i=6}^{9}
(\partial_{\tilde r}X^i)^2+ h_2^{-1}(r)(\partial_{\tilde r}X^{10})^2\right)\\
B(r) &=& \sqrt{-h_2^{1/3}(r)\sum_{i=6}^{9} (\partial_{\tilde r}X^i)^2}
\end{eqnarray}
We have put a tilde on $H$ to distinguish it from that obtained in the flat
background. We will continue to use the same notation in the {\it M5} brane
background as well. The BPS condition gives rise to a quadratic equation for
$\tilde H_{1\tilde r}$. We, however, end up with only one solution because the
term inside the square root of the solution is negative semi-definite and a
real solution for $\tilde H_{1\tilde r}$ is obtained by setting that term to
zero. This is achieved by setting $\partial_{\tilde r}X^{i}=0$ for $i=6,7,8,9$.
Therefore,
\begin{equation}\label{hm2}
\tilde H_{1\tilde r} = -h_2^{-2/3}(r)\partial_{\tilde r}X^{10}
\sqrt{1+ h_2^{-1}(r)(\partial_{\tilde r}X^{10})^2}.
\end{equation}
This solution should also satisfy the Bianchi identity
constraint (\ref{bic}), which in this background (\ref{m2onm5})
becomes
\begin{equation}\label{curvhm2}
\partial_{\tilde r}(\tilde r^3 h_2^{1/3}(r)\partial_{\tilde r}X^{10})=0.
\end{equation}
It is interesting to see that for large values of $\tilde r$, the
Bianchi identity leads to same equation for $X_{10}$ even in the
case of a curved background. Hence the solution to this equation is
same as that given in eq.(\ref{soln}) in this limit. The global
solution, however, differs from the flat case. This can be seen,
{\it e.g.}, in the near horizon limit where we can ignore constant
element in the harmonic function $h_2(r)$ and taking $r=\tilde r$
limit,
\begin{equation}\label{bim2x}
X_{10}(\tilde r) = \frac{c_2}{N_2^{1/3}\ell_p^2}\ln\tilde r.
\end{equation}
The fact that in the large $\tilde r$ limit, the Bianchi identity
reduces to that in the flat space is not surprising and can be used
to determine the charge carried by the string soliton. The charge
$N_{2s}$ carried by the string soliton is determined by integrating
$\tilde H_{1\tilde r}$ over the asymptotic $S^3$ which encloses the
soliton. Using this condition we can determine asymptotic behaviour
of $X_{10}$ for large values of $\tilde r$ and we get
\begin{equation}\label{m2solch}
X_{10}(\tilde r) \sim \frac{N_{2s}}{\tilde r^2}.
\end{equation}

Now we will look at the worldvolume string soliton in the {\it M5}
brane background (\ref{m5onm5}). We determine the three form field
configuration which solves the BPS condition (\ref{bpsc}) for the
string soliton by using the induced metric (\ref{m5onm5}). The field
strength $\tilde H_{1\tilde r}$ is
\begin{equation}\label{hm5}
\tilde H_{1\tilde r} = - h_5^{1/6}(r)\sqrt{1+(\partial_{\tilde
r}X^{10})^2}(\partial_{\tilde r}X^{10}).
\end{equation}
Like in the {\it M2} brane background, scalar fields $X^i$ for
$i=6,7,8,9$ are set to zero due reality condition on the field
strength $\tilde H_{1\tilde r}$. The Bianchi identity gives us
\begin{equation}\label{bim5}
\partial_{\tilde r}(\tilde r^3 h_5^{7/6}(r)\partial_{\tilde r}X^{10})=0.
\end{equation}
Notice, in the large $\tilde r$, this equation is same as that
obtained in the {\it M2} brane background. However, for finite
values of $\tilde r$, this equation differs significantly from that
obtained in the {\it M2} brane background. Particularly, in the near
horizon limit and with $r =\tilde r$, we can determine behaviour of
$X_{10}(\tilde r)$ using the Bianchi identity constraint,
\begin{equation}\label{bicm5sx}
X_{10}(\tilde r) = \frac{c_5}{N_5^{7/6}\ell_p^{7/2}}\tilde r^{3/2}.
\end{equation}
This behaviour is significantly different from that obtained in the
{\it M2} brane background. Large $\tilde r$ behaviour of
$X_{10}(\tilde r)$ is again determined by either using the Bianchi
identity or using the Gauss' law constraint. The latter determines
behaviour of $X_{10}(\tilde r)$ as a function of soliton charge
$N_{5s}$ for large $\tilde r$,
\begin{equation}\label{solch5}
X_{10}(\tilde r) \sim \frac{N_{5s}}{\tilde r^2}.
\end{equation}

\section{Conclusions and discussion}

The motivation behind this paper was to investigate the claim \cite{gaunt}
that two M5-branes can never intersect along one direction without a membrane
being present; when it can not be seen, one is to assume  that the membrane
has collapsed. Building on \cite{braneteaser}, (which stated that two
M5-branes {\it can} in fact intersect along a string if the world-volume
two-form is turned on), we study one-dimensional intersections of M5-branes.
 In particular, we focus on $M5\perp M5(1)$ and $M5\perp M2(1)$
intersections, studying these from the point of view of the
fivebrane worldvolume.

$M5\perp M5(1)$ and $M5\perp M2(1)$
intersections preserve the same fraction of supersymmetry in addition
to having the same isometries. In the worldvolume theory of an
M5-brane, the string solitons corresponding to each of these one
dimensional intersections are explored.

We considered, in turn, both M-brane intersections as worldvolume
solitons in a probe M5-brane.  We were lead to the same mathematical
form of the solution for the soliton.  Even though the solutions
have the same form one can ask whether they are in fact identical.
The question thus is what the worldvolume charge produced by each
one of the intersections is.  Even though we do not fully answer
this question in this setup, we do find evidence that the charges
should be different in a slightly different context later.

In our second approach to the problem we look at a probe M5-brane in
a curved space-time produced, respectively, by stacks of M2 and
M5-branes.  In both cases the worldvolume $H$ field vanishes due to
local canceling of charges as the intersecting brane "passes right
through" the probe M5-brane.   We compare the induced worldvolume
metrics on the probe M5-brane by taking the Maldacena decoupling
limit in which the induced metrics are both $AdS_3\times S^3$ but
with differing radii of curvature.  Giving us some evidence that the
M2 and M5-branes do indeed appear in a distinguishable way in the
worldvolume theory.  Finally, we find the most general string
soliton solutions in the worldvolume theory in the M2 and M5 brane
background.

\vspace{5mm}

{\bf Acknowledgements}:\\

We would like to thank the Physics Department at Harvard University,
and in particular, Shiraz Minwalla and Cumrun Vafa for their warm
hospitality. This work is supported in part by NSF Award No. PHY
0239626, Department of Energy Award No. DE-FG02-ER9140654 and Sloan
Award No. BR-4142. AF is supported by a grant from
Vetenskapsr{\aa}det.

\end{document}